\documentclass[preprint,12pt]{elsarticle}
\usepackage{graphics}
\usepackage{mathrsfs}
\usepackage{amsmath}

\journal{Physica A}

\begin{document}

\begin{frontmatter}

\title{Order formation processes of complex systems \\ including different parity order parameters}


\author[1]{Yoichiro Hashizume}
\address[1]{Tokyo University of Science,\\ 6-3-1 Niijuku, Katsushika-ku, Tokyo, 125-8585, Japan}
\ead{hashizume@rs.tus.ac.jp}
\author[2]{Masuo Suzuki}
\address[2]{Computational Astrophysics Laboratory, RIKEN,\\ 2-1 Hirosawa, Wako, Saitama, 351-0198, Japan}
\ead{masuo.suzuki@riken.jp}
\author[1]{Soichiro Okamura}
\ead{sokamura@rs.kagu.tus.ac.jp}

\begin{abstract}
In the present study, we focus on the parity of the order parameters and clarify the order formation process in a system including two order parameters.
Each order parameter shows different parity under a gauge transformation, namely even and odd order parameters.
For example, in a spin-glass model, the even order parameter corresponds to the spin-glass order parameter while the odd one corresponds to the magnetization.
We introduce phenomenologically a set of Langevin equations to express the ordering process under a white Gaussian noise.
Using two kinds of Fokker-Planck equations, we analyze the order formation process and the entropy production.
Furthermore, we show the noise dependence of the onset time.
\end{abstract}


\end{frontmatter}

\section{Introduction}
\label{intro}

The scaling theory of order formation processes was established by one of the authors (M.S.) in 1976 \cite{1,2}.
This scaling theory focuses on an order parameter, namely magnetization, and clarifies the process from an unstable state to the final stable state.
According to the scaling theory \cite{1,2}, the order formation process can be expressed by the asymptotic expansion with respect to the scaling time $\tau$, namely $\tau \propto e^{2\gamma t}$, where $t$ denotes time, and $\gamma$ a constant real number \cite{1,2}.
The scaling theory can be applied to many phenomena including fish schooling \cite{3} and nuclear fission \cite{4}.
In parallel to the works on applications, it was repeatedly tested [5-7] using analytic and numerical methods.

Recently, relaxation processes including the order formation become more important in such complex systems as glass-like systems [8-10].
In many complex systems, there exists a highly symmetric structure in an apparently disordered state.
For example, in the spin-glass phase, no magnetization appears, while a spin-glass ordered state appears.
In the present study, we clarify the ordering processes in a system including two order parameters, namely even-parity order-parameters (namely, ``even order parameters") and odd-parity order-parameters (namely, ``odd order parameters").
This is one of the typical examples in the above complex systems.

At first, we introduce a set of Langevin equations describing our system with even and odd order parameters.
Thus we analyze directly the Langevin equations in Section 2.
Using the distribution functions, the Langevin equations can be represented by the Fokker-Planck equations.
We solve the Fokker-Planck equations and obtain the relevant onset time in our system in Section 3.
The time evolution of the entropy is obtained from the distribution function as shown in Section 4.
A typical example of the present order formation process appears in the four-body interaction model \cite{9,10} and a spin-glass model. 
We discuss a simple derivation of this process from the four-body interaction model in Section 5.
A summary and discussions are given in Section 6.

\section{Langevin equations including even and odd order parameters}
\subsection{Basic theory of order formation}
In this subsection, we make a brief review of the basic theory of the order formation \cite{1,2} using our notation.
It may be useful to introduce our Langevin equations.
The basic theory \cite{1,2} expresses the phase transition from an initial unstable state to the final stable state.
The time dependence of the order parameter $x(t)$ on the double well potential $V(x)=-(\gamma/2)x^2+(g/4)x^4$ is obtained by the Langevin equation \cite{1,2}
\begin{equation}
\frac{d}{dt}x(t)=\gamma x(t)-gx^3(t)+\zeta(t),\label{eq1}
\end{equation}
where the parameters $\gamma$ and $g$ depend on the temperature $T$.
Especially the parameter $\gamma$ takes a positive value ($\gamma >0$) for $T<T_{\text{c}}$, where the temperature $T_{\text{c}}$ denotes the critical temperature.
Note that the noise $\zeta(t)$ is assumed to be a white Gaussian noise:
\begin{equation}
\langle \zeta(t)\zeta(t')\rangle = 2 \epsilon \delta (t-t'), \label{eq2}
\end{equation} 
where the average $\langle Q(t) \rangle$ denotes the stochastic average of the random parameter $Q(t)$ under the random noise $\zeta (t)$, and the constant value $\epsilon$ denotes the strength of the noise.

Using the Langevin equation (\ref{eq1}), the fluctuation $\langle x^2 \rangle_t$ of the order parameter $x(t)$ is obtained  as \cite{1,2}
\begin{equation}
\langle x^2 \rangle_t \simeq \frac{\langle x^2 \rangle_{\infty}}{\sqrt{2\pi}}\int_{-\infty}^{\infty}\frac{\xi^2\tau}{1+\xi^2 \tau}e^{-\xi^2/2}d\xi \label{eq3}
\end{equation}
under the scaling limit \cite{1,2}
\begin{equation}
\epsilon \to 0, g\to 0, t\to \infty {\text{ and }}\tau={\text{finite}}, \label{eq4}
\end{equation}
where the scaling time $\tau$ is defined \cite{1,2} by
\begin{equation}
\tau \equiv \frac{g\epsilon}{\gamma^2}e^{2\gamma t}.\label{eq5}
\end{equation}
The scaling limit (\ref{eq4}) means the limiting-case of the small noise or small nonlinearity.
This scaling limit corresponds to the condition which appears in the order formation processes.
As shown in Eqs.(\ref{eq3}) and (\ref{eq5}), the order formation process is scaled exponentially by the scaling time $\tau$.
According to this scaling theory, an ordered state appears in the following time scale, namely ``onset time" \cite{1,2}:
\begin{equation}
O(\tau)=1\Leftrightarrow t_{\text{o}}=\frac{1}{2\gamma}\log\frac{\gamma^2}{g\epsilon}.\label{eq6}
\end{equation}

\subsection{Classification of order parameters in terms of symmetry}
\begin{figure}[hbpt]
\begin{center}
\includegraphics[width=7cm]{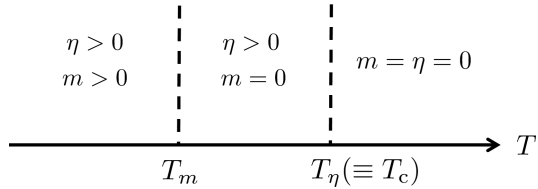}
\end{center}
\caption{Typical phase diagram. In many cases, the lower transition point $T_m$ is zero. The phase characterized by the conditions $\eta>0$ and $m=0$ appears in the region $T_m<T<T_{\eta}$.}
\label{fig1}
\end{figure}

As discussed in Section 1, in many complex systems, there exists a highly symmetric structure in an apparently disordered state.
Then we classify the order parameters from the view point of the parity, that is, one is the odd-parity order-parameter (namely, the ``odd order parameter") and another is the even-parity order-parameter (namely, the ``even order parameter").
A typical example including these order parameters is shown in four-body interaction models \cite{9,10}.
In this example, the magnetization corresponds to the {\it odd} order parameter, while the spin-pair corresponds to the {\it even} order parameter.
In four-body interaction models, under the gauge transformation $\{S_i\}\to \{-S_i\}$, the magnetization $m\equiv \langle S_i \rangle$ changes into $-m$, while the spin-pair order parameter $\eta\equiv \langle S_iS_j \rangle$ does not change.
Then, in general, the even order parameter is denoted by $\eta$ and the odd order parameter is denoted by $m$ below.
Here the two characteristic temperatures $T_{\eta}$ and $T_{m}$ can be defined as the transition temperatures corresponding to the even and odd order parameters, respectively, as shown in Fig.\ref{fig1}.
Because the even order parameter $\eta$ has higher symmetric property, the transition temperature $T_{\eta}$ is higher than $T_{m}$.
In many cases, such as four-body interaction models or some spin-glass models, the lower critical temperature $T_m$ is zero.
Then we focus on the phase in the temperatures $T_m<T<T_{\eta}$.
Thus, we may conclude $T_{\text{c}}$ denotes the critical temperature $T_{\text{c}}$ of the system.

\subsection{Langevin equations including the two order parameters $m$ and $\eta$}
In general, the Langevin equations including odd and even order parameters are shown as
\begin{equation}
\tau_{m}\frac{d}{dt}m=F(m,\eta)+\zeta(t) \label{eq7}
\end{equation}
and
\begin{equation}
\tau_{\eta}\frac{d}{dt}\eta =G(m,\eta)+\zeta'(t).\label{eq8}
\end{equation}
Here the parameters $\tau_{m}$ and $\tau_{\eta}$ denote the time constants of $m$ and $\eta$, and the white Gaussian noises $\zeta(t)$ and $\zeta'(t)$ satisfy the conditions
\begin{equation}
\langle \zeta(t)\zeta(t') \rangle = 2\epsilon \delta (t-t') {\text{ and }} \langle \zeta'(t)\zeta'(t') \rangle = 2\epsilon' \delta (t-t'). \label{eq9}
\end{equation}
When we focus on the region $T_m<T<T_{\eta}$, the odd order parameter $m$ vanishes, while the even order parameter $\eta$ becomes a positive finite value (namely, $\eta>0$) as shown in Fig.\ref{fig1}.
Phenomenologically, since the parameter $m$ does not order at any temperature, the dynamics of $m$ is described by the harmonic potential.
Then the force $F(m,\eta)$ in Eq.(\ref{eq7}) is defined as
\begin{equation}
F(m,\eta)\equiv -\frac{\partial}{\partial m} V_{F}(m);\quad V_{F}(m)\equiv \frac{1}{2}m^2. \label{eq10}
\end{equation}
On the other hand, the ordered state characterized by the parameter $\eta$ is induced by the fluctuation of the parameter $m$.
Thus the potential $V_{G}(\eta)$ is assumed to be a double well potential.
Then we introduce the potential the force $G(m,\eta)$ of the form
\begin{equation}
G(m,\eta)\equiv -\frac{\partial}{\partial \eta} V_{G}(\eta)+cm^2\eta;\quad V_{G}(\eta)\equiv -\frac{a}{2}\eta^2+\frac{b}{4}\eta^4,\label{eq11}
\end{equation}
where the parameters $a,b$ and $c$ depend on the temperatures as follows:
\begin{equation}
a=a_0(T_{\text{c}}-T),\quad b=b(T)>0,\quad c=c(T)>0.\label{eq12}
\end{equation}
Here, we treat the case that the direct noise can be ignored $\epsilon'=0$ but the minimal coupling effect $cm^2\eta$ is important.
Finally, we obtain the Langevin equations
\begin{equation}
\tau_{m}\frac{d}{dt}m=-m+\zeta(t) \label{eq13}
\end{equation}
and
\begin{equation}
\tau_{\eta}\frac{d}{dt}\eta=a\eta-b\eta^3+cm^2\eta \label{eq14}
\end{equation}
with respect to the order parameters $m$ and $\eta$.

\begin{figure}[hbpt]
\begin{center}
\includegraphics[width=7cm]{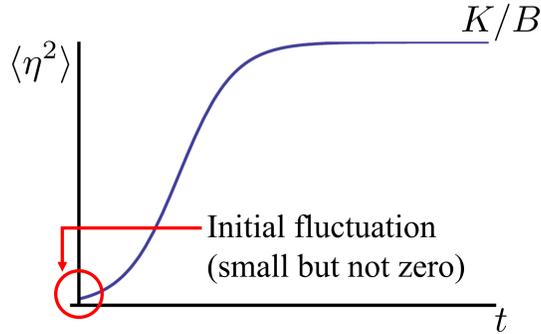}
\end{center}
\caption{Time evolution of the fluctuation $\langle \eta^2 \rangle$.}
\label{fig2}
\end{figure}

In the scaling region, the time evolution of the fluctuation $\langle \eta^2\rangle_{t}$ of the parameter $\eta$ is more rapid than that of the parameter $m$, namely $\langle m^2 \rangle _{t}$.
Then we assume that the condition $\tau_{m}\gg \tau_{\eta}$ is satisfied and that $\langle m^2 \rangle$ takes a constant value.
Thus we arrive at the expression of the fluctuation of the even order parameter $\eta$ as
\begin{equation}
\langle \eta^2 \rangle =\frac{\langle \eta^2 \rangle_0 \tilde{\tau}^2}{1+\langle \eta^2 \rangle_0 \tilde{\tau}^2}\langle \eta^2 \rangle_{\infty} \label{eq15}
\end{equation}
from Eq.(\ref{eq14}), using the initial fluctuation $\langle \eta^2 \rangle_0$ and the stable state value $\langle \eta^2 \rangle_{\infty}=K/B$.
Here the parameters $K$ and $B$ are defined as $K=(a+c\langle m^2 \rangle)/\tau_{\eta}$ and $B=b/\tau_{\eta}$, respectively.
This is because the equation (\ref{eq14}) yields
\begin{equation}
\frac{1}{2}\frac{d\langle \eta^2 \rangle}{dt}=K\langle \eta^2 \rangle - B\langle \eta^4 \rangle\simeq K\langle \eta^2 \rangle - B\langle \eta^2 \rangle^2,\label{eq16}
\end{equation}
using the decoupling approximation.
Here the scaling time $\tilde{\tau}$ is defined as
\begin{equation}
\tilde{\tau}\equiv \sqrt{\frac{B}{K}}e^{Kt}.\label{eq17}
\end{equation}
The scaling form (\ref{eq15}) includes the initial fluctuation $\langle \eta^2 \rangle_0$ of the order parameter $\eta$.
If it takes the initial condition $\langle \eta^2 \rangle_0 =0$, the ordered state does not appear.
It corresponds to the situation that the parameter $\eta$ is not affected directly by the noise $\zeta(t)$.
Then the ordered state characterized by the even order parameter $\eta$ is induced even by very small (but not zero) initial fluctuations as shown in Fig.\ref{fig2}.

It is interesting to note that the fluctuation (\ref{eq15}) is similar to the kernel of the fluctuation (\ref{eq3}).
However, the fluctuation (\ref{eq15}) is not expressed by the Gaussian weighted integral (\ref{eq3}).
This difference is based on the types of noises.
Contrary to the traditional scaling theory \cite{1}, the present Langevin equations (\ref{eq13}) and (\ref{eq14}) include a ``multiplicative" noise.
Additionally, we have assumed conditions $\tau_{m}\gg \tau_{\eta}$ and $\langle m^2 \rangle (t\sim t_{\text{o}})={\text{constant}}$.
This assumption corresponds to the scaling limit (\ref{eq4}) and yields such a simple expression as shown in Eq.(\ref{eq15}).

\section{Fokker-Planck approach --distribution function and onset time--}
The Langevin equations (\ref{eq13}) and (\ref{eq14}) are represented by the Fokker-Planck equations \cite{11,12} as
\begin{equation}
\frac{\partial}{\partial t}P^{(m)}(t,m)=\frac{1}{\tau_m}\frac{\partial}{\partial m}m P^{(m)}(t,m)+\frac{\epsilon}{\tau_m^2}\frac{\partial^2}{\partial m^2}P^{(m)}(t,m)\label{eq18}
\end{equation}
and
\begin{equation}
\frac{\partial}{\partial t}P^{(\eta)}(t,\eta,m)=\frac{\partial}{\partial \eta}\alpha(\eta)P^{(\eta)}(t,\eta,m).\label{eq19}
\end{equation}
Here the function $\alpha(\eta)$ denotes
\begin{equation}
\alpha(\eta)\equiv \frac{1}{\tau_{\eta}}\left[(a+cm^2)\eta-b\eta^3\right] \equiv K(m)\eta-B\eta^3.\label{eq20}
\end{equation}
From the discussion in Section 2, we find  that it is necessary to include initial fluctuations of the order parameter $\eta$.
Then the initial conditions are assumed as follows:
\begin{equation}
P^{(m)}(0,m)=\frac{1}{\sqrt{2\pi\sigma_{m}^2}}\exp\left[-\frac{m^2}{2\sigma_{m}^2}\right]\label{eq21}
\end{equation}
and
\begin{equation}
P^{(\eta)}(0,\eta,0)=\frac{1}{\sqrt{2\pi\sigma_{\eta}^2}}\exp\left[-\frac{\eta^2}{2\sigma_{\eta}^2}\right]\equiv f(\eta),\label{eq22}
\end{equation}
where the variances $\sigma_{m}$ and $\sigma_{\eta}$ mean the initial fluctuations of the parameters $m$ and $\eta$, respectively.
As has been discussed in the previous section, we focus on the scaling region of the parameter $\eta$.
In this scaling region, the time evolution of the fluctuation $\langle m^2 \rangle$ is slower than that of $\eta$.
Then the parameter $K(m)$ can be assumed to take a constant value and the initial fluctuations $\sigma_{m}$ and $\sigma_{\eta}$ satisfy the condition $\sigma_{m}\gg \sigma_{\eta}$.  

The distribution functions $P^{(m)}(t,m)$ and $P^{(\eta)}(t,\eta,m)$ can be obtained explicitly using the characteristic curve method as
\begin{equation}
P^{(m)}(t,m)=\frac{1}{\sqrt{2\pi \left(\sigma_{m}^2+(\epsilon/\tau_m) (1-e^{-2t/\tau_m})\right)}}\exp\left[-\frac{m^2}{2 \left(\sigma_{m}^2+(\epsilon/\tau_m) (1-e^{-2t/\tau_m})\right)}\right]\label{eq23}
\end{equation}
and
\begin{align}
P^{(\eta)}(t,\eta,m)=e^{-K(m)t} &\left[ \frac{K(m)}{K(m)+(e^{-2K(m)t}-1)B\eta^2} \right]^{3/2} \notag
\\
& \times f\left(\left[ \frac{K(m)e^{-2K(m)t}\eta^2}{K(m)+(e^{-2K(m)t}-1)B\eta^2} \right]^{1/2}\right).\label{eq24}
\end{align}
These expressions are too complicated to derive explicitly here.
However, it is easily confirmed that they actually satisfy the initial conditions (\ref{eq21}) and (\ref{eq22}), and the partial differential equations (\ref{eq18}) and (\ref{eq19}).

The distribution functions (\ref{eq23}) and (\ref{eq24}) show a typical behavior of the order formation processes with plural order parameters.
The distribution function $P^{(m)}(t,m)$ has a peak at $m=0$ in the whole region of time.
Consequently, we can understand that the ordered state characterized by the odd order parameter $m$ does not appear in any time.
On the other hand, the distribution function $P^{(\eta)}(t,\eta,m)$ shows double peaks around the stable states $\eta=\pm \eta_{\text{eq}}=\pm \sqrt{K(m)/B}$ for larger $t$.
Then, we can understand that the ordered state characterized by the even order parameter $\eta$ appears in the equilibrium state.
These typical behaviors generally appear in the order formation process on a complex system characterized by plural order parameters.

\begin{figure}[hbpt]
\begin{center}
\includegraphics[width=12cm]{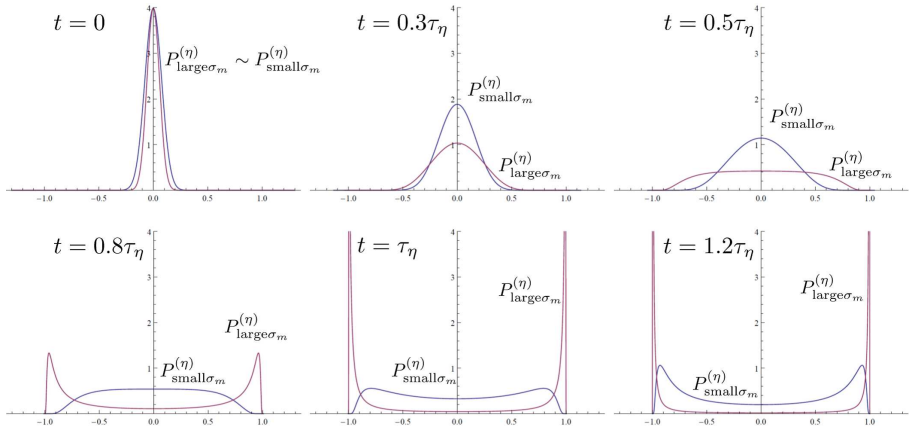}
\end{center}
\caption{$\sigma_{m}$ dependence of the distribution functions of the order parameter $\eta$, namely $P^{(\eta)}_{{\text{large}}\sigma_{m}}$ and $P^{(\eta)}_{{\text{small}}\sigma_{m}}$. The horizontal axis shows the order parameter $\eta$ while the vertical axis shows the distribution functions. The distribution function of the order parameter $\eta$, $P^{(\eta)}_{{\text{large}}\sigma_{m}}$, corresponds to such a large $\sigma_{m}$ as satisfies the equation $a+c\sigma_{m}^2=4.5$, while $P^{(\eta)}_{{\text{small}}\sigma_{m}}$ corresponds to such a small $\sigma_{m}$ as satisfies the equation $a+c\sigma_{m}^2=2.5$. The conditions of the parameters $B$ and $\sigma_{\eta}$ are assumed to be $B=1.5$ and $\sigma_{\eta}=0.1$. We can clearly find that the larger fluctuation of the odd order parameter $m$ induces the order formation of the even order parameter $\eta$ as shown in Eq.(\ref{eq30}). }
\label{fig3}
\end{figure}

Now, we consider the onset time of the even order parameter $\eta$.
The distribution function $P^{(\eta)}(t,\eta,m)$ in Eq.(\ref{eq24}) can be represented as
\begin{equation}
P^{(\eta)}(t,\eta,m)=e^{-K(m)t}\left[ \frac{1}{1-\tau'\eta^2} \right]^{3/2}f\left(\left[ \frac{e^{-2K(m)t}\eta^2}{1-\tau'\eta^2} \right]^{1/2}\right),\label{eq25}
\end{equation}
using the parameter $\tau'$ defined by
\begin{equation}
\tau'\equiv \frac{B}{K(m)}\left( 1-e^{-2K(m) t} \right).\label{eq26}
\end{equation}
This parameter $\tau'$ corresponds to the scaling time $\tilde{\tau}$ defined in Eq.(17) as
\begin{equation}
e^{2K(m)t}\tau'=\frac{B}{K(m)}\left( e^{2K(m) t}-1 \right)\simeq \left(\sqrt{\frac{B}{K(m)}}e^{K(m)t}\right)^2=\tilde{\tau}^2\label{eq27}
\end{equation}
for larger $t$.
We assume again the condition $\sigma_m\gg \sigma_{\eta}$.
In this case, we obtain
\begin{equation}
K(m)\simeq \frac{1}{\tau_{\eta}}(a+c\langle m^2 \rangle)\simeq \frac{1}{\tau_{\eta}}(a+c \sigma_{m}^2)\label{eq28}
\end{equation}
for larger $t$, because the fluctuation $\langle m^2 \rangle$ of the odd order parameter $m$ is obtained as
\begin{equation}
\langle m^2 \rangle =\sigma_{m}^2 + \frac{\epsilon}{\tau_m}\left(1-e^{-2t/\tau_m}\right)\to \sigma_{m}^2 + \frac{\epsilon}{\tau_m}\to \sigma_{m}^2 \label{eq29}
\end{equation}
from the distribution function (\ref{eq23}) for larger $t$.
The onset time $t_{\text{o}}^*$ is defined by the condition $O(\tilde{\tau})\sim 1$ \cite{1,2}.
Then the onset time $t_{\text{o}}^*$ is derived as
\begin{equation}
\sqrt{\frac{B}{K(m)}}e^{K(m)t_{\text{o}}^*}=1\Leftrightarrow t_{\text{o}}^*=\frac{\tau_{\eta}}{2(a+c\sigma_{m}^2)}\log\frac{a+c\sigma_{m}^2}{b}\label{eq30}
\end{equation}
using the relations (\ref{eq27}) and (\ref{eq28}).
By the way, the traditional onset time $t_{\text{o}}$ is shown \cite{1,2} in Eq.(\ref{eq6}).
Comparing the onset time $t_{\text{o}}^*$ with $t_{\text{o}}$, they are essentially different in the noise dependence.
As shown in Eq.(\ref{eq6}), the onset time $t_{\text{o}}$ of the {\it additive} noise system (\ref{eq1}) depends logarithmically on the noise intensity $\epsilon$ as $t_{\text{o}}\sim \log \epsilon^{-1}$.
On the other hand, in our {\it multiplicative} system, the onset time $t_{\text{o}}^*$ depends on the fluctuation $\sigma_{m}$ as $t_{\text{o}}^*\sim \sigma_{m}^{-2}$.
Thus, we can conclude that the fluctuation $\sigma_{m}^2$ of the odd order parameter $m$ is essentially important for the order formation of the even order parameter $\eta$ as shown in Fig.{\ref{eq3}}.

\section{Entropy production}
In this section, we discuss the entropy production using the distribution functions $P^{(m)}(t,m)$ and $P^{(\eta)}(t,\eta,m)$ as shown in Eqs.(\ref{eq23}) and (\ref{eq24}).
The entropy of non-equilibrium systems was recently discussed by one of the authors (M.S.) \cite{13}.
The entropy $S^{(\Omega)}(t)$ is defined as
\begin{equation}
S^{(\Omega)}(t)=-k_{\text{B}}\int_{\Omega}P^{(\Omega)}(t)\log P^{(\Omega)}_{\text{eq}}d\Omega,\label{eq31}
\end{equation}
where the notation $\Omega$ means the corresponding order parameter, namely $m$ or $\eta$.
The distribution function in equilibrium, $P^{(\Omega)}_{\text{eq}}$, is defined in the limit $t\to \infty$, namely
\begin{equation}
P^{(\Omega)}_{\text{eq}}=\lim_{t\to\infty}P^{(\Omega)}(t).\label{eq32}
\end{equation}
For example, for the odd order parameter $m$, the distribution function $P^{(m)}_{\text{eq}}$ is derived as
\begin{align}
P^{(m)}(t,m)&=\frac{1}{\sqrt{2\pi \left(\sigma_{m}^2+(\epsilon/\tau_m) (1-e^{-2t/\tau_m})\right)}}\exp\left[-\frac{m^2}{2 \left(\sigma_{m}^2+(\epsilon/\tau_m) (1-e^{-2t/\tau_m})\right)}\right]\notag
\\
&\to \frac{1}{\sqrt{2\pi \left(\sigma_{m}^2+\epsilon/\tau_m \right)}}\exp\left[-\frac{m^2}{2 \left(\sigma_{m}^2+\epsilon/\tau_m\right)}\right]
\equiv P^{(m)}_{\text{eq}}.\label{eq33}
\end{align}
Then the entropy $S^{(m)}(t)$ corresponding to the parameter $m$ is obtained as
\begin{align}
S^{(m)}(t)&=-k_{\text{B}}\int_{-\infty}^{\infty} P^{(m)}(t,m)\log  P^{(m)}_{\text{eq}} dm \notag
\\
&=-k_{\text{B}}\langle \log  P^{(m)}_{\text{eq}} \rangle \notag
\\
&=\frac{k_{\text{B}}}{2\left(\sigma_{m}^2+ \epsilon/\tau_m \right)} \langle m^2 \rangle +\frac{k_{\text{B}}}{2}\log {2\pi \left(\sigma_{m}^2+ \epsilon/\tau_m \right)} \notag
\\
&=\frac{k_{\text{B}}}{2}\left(1-\frac{\epsilon /\tau_{m}}{\sigma_{m}^2+\epsilon/\tau_{m}}e^{-2t/\tau_{m}}\right)+\frac{k_{\text{B}}}{2}\log 2\pi (\sigma_{m}^2+\epsilon/\tau_{m})
\label{eq34}
\end{align}
As seen in Eq.(\ref{eq34}), the entropy  $S^{(m)}(t)$ corresponds to the fluctuation $\langle m^2 \rangle$ in this system.
Furthermore, the entropy production $\sigma^{(\Omega)}(t)\equiv \partial S^{(\Omega)}(t)/\partial t$ is obtained as
\begin{equation}
\sigma^{(m)}(t)=\frac{\partial S^{(m)}(t)}{\partial t}=\frac{k_{\text{B}}}{\tau_{m}}\frac{\epsilon /\tau_{m}}{\sigma_{m}^2+\epsilon/\tau_{m}}e^{-2t/\tau_m}.\label{eq35}
\end{equation}
Clearly, the entropy production $\sigma^{(m)}(t)$ is positive.
Then the entropy $S^{(m)}(t)$ increases.
The time evolution of the distribution function of the odd parameter $m$ corresponds to the above results.

\begin{figure}[hbpt]
\begin{center}
\includegraphics[width=10cm]{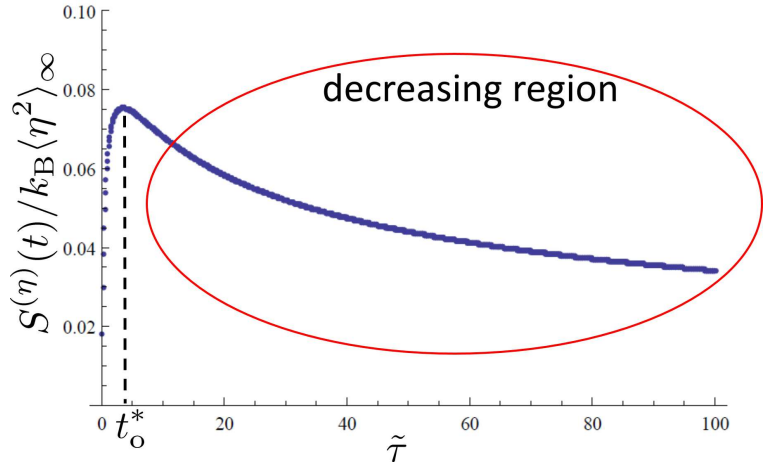}
\end{center}
\caption{Time dependence of the entropy $S^{(\eta)}(t)$. The variance $\sigma_{\eta}^2$ is assumed as $\sigma_{\eta}^2=0.8$. We can see the entropy decreasing in larger $t$.}
\label{fig4}
\end{figure}

On the other hand, from the distribution function (\ref{eq24}) of the even order parameter $\eta$, the asymptotic expression of the equilibrium distribution function $P^{(\eta)}_{\text{eq}}$ is derived as follows:
\begin{align}
P^{(\eta)}(t,\eta,m) =&e^{-K(m)t}\left[ \frac{K(m)}{K(m)+(e^{-2K(m)t}-1)B\eta^2} \right]^{3/2}\notag
\\
&\times f\left(\left[ \frac{K(m)e^{-2K(m)t}\eta^2}{K(m)+(e^{-2K(m)t}-1)B\eta^2} \right]^{1/2}\right) \notag
\\
=&\frac{1}{\sqrt{2\pi\sigma_{\eta}^2}}\left( \frac{\langle \eta^2 \rangle_{\infty}}{\langle \eta^2 \rangle_{\infty}-\eta^2} \right)^{3/2}\notag
\\
&\times \exp \left[ -K(m)t -\frac{1}{2\sigma_{\eta}^2}\frac{\eta^2}{1-\eta^2/\langle \eta^2 \rangle_{\infty}} e^{-2K(m)t} \right] \notag
\\
\simeq &\frac{1}{\sqrt{2\pi\sigma_{\eta}^2}}\left( \frac{\langle \eta^2 \rangle_{\infty}}{\langle \eta^2 \rangle_{\infty}-\eta^2} \right) ^{3/2}e^{-K(m)t} \label{eq36}
\end{align}
for the condition $t\to \infty$.
Here we assume the initial condition $f(x)$ to a Gaussian distribution as shown in Eq.(\ref{eq22}).
From the asymptotic expression (\ref{eq36}), we find that the distribution function $P^{(\eta)}_{\text{eq}}$ has double peaks around the stable point $\eta\sim \pm \langle |\eta| \rangle_{\infty}=\pm \sqrt{K(m)/B}$ for larger $t$.
According to this discussion, the distribution function $P^{(\eta)}_{\text{eq}}$ can be estimated as
\begin{equation}
P^{(\eta)}_{\text{eq}} \sim \left\{
\begin{array}{l}
\exp\left[-\frac{1}{2\sigma_{\eta}^2}(\eta-\langle |\eta|\rangle)^2\right] \quad ({\text{for }}\eta>0)
\\
\\
\exp\left[-\frac{1}{2\sigma_{\eta}^2}(\eta+\langle |\eta|\rangle)^2\right] \quad ({\text{for }}\eta<0).
\end{array}
\right.
\label{eq37}
\end{equation}
For this $P^{(\eta)}_{\text{eq}}$, the entropy $S^{(\eta)}(t)$ is obtained as
\begin{align}
S^{(\eta)}(t)=&-k_{\text{B}}\int_{-\infty}^{\infty} P^{(\eta)}(t,\eta, m)\log  P^{(\eta)}_{\text{eq}} d\eta \notag
\\
=&-k_{\text{B}}\int_{-\infty}^{0} P^{(\eta)}(t,\eta, m)\frac{-1}{2\sigma_{\eta}^2}(\eta+\langle |\eta|\rangle)^2d\eta \notag
\\
&- k_{\text{B}}\int_{0}^{\infty} P^{(\eta)}(t,\eta, m)\frac{-1}{2\sigma_{\eta}^2}(\eta - \langle |\eta|\rangle)^2d\eta\notag
\\
\simeq &2k_{\text{B}}\int_{0}^{\infty} P^{(\eta)}(t,\eta, m)(\eta - \langle |\eta|\rangle)^2d\eta \notag
\\
=&k_{\text{B}}\left( \langle \eta^2 \rangle -\langle |\eta| \rangle^2 \right).\label{eq38}
\end{align}
Unfortunately, the fluctuation in Eq.(\ref{eq38}) is too complicated to express explicitly.
Then we calculate the entropy $S^{(\eta)}(t)$ numerically and show the results in Fig.\ref{fig4}.
Here the numerical calculation is performed using the relations
\begin{align}
\langle |\eta|\rangle &= 2\int_{0}^{\infty}\eta P^{(\eta)}(t,\eta, m)d\eta \notag
\\
&=\langle |\eta| \rangle_{\infty}\times 2\tilde{\tau}^{1/2}\int_0^{\sqrt{1/\tilde{\tau}}} x \left( \frac{1}{1-\tilde{\tau}x} \right)^{3/2}f\left(\left[ \frac{x^2}{1-\tilde{\tau}x} \right]^{1/2} \right)dx, \label{eq39}
\end{align}
\begin{align}
\langle \eta^2 \rangle &= \int_{-\infty}^{\infty}\eta^2 P^{(\eta)}(t,\eta, m)d\eta \notag
\\
&=\langle |\eta| \rangle_{\infty}\times \tilde{\tau}\int_{-\sqrt{1/\tilde{\tau}}}^{\sqrt{1/\tilde{\tau}}} x^2 \left( \frac{1}{1-\tilde{\tau}x} \right)^{3/2}f\left(\left[ \frac{x^2}{1-\tilde{\tau}x} \right]^{1/2} \right)dx, \label{eq40}
\end{align}
and
\begin{equation}
\langle |\eta| \rangle^2_{\infty}=\langle \eta^2 \rangle_{\infty}=\frac{K(m)}{B}\simeq \frac{a-c\langle m^2 \rangle}{b}. \label{eq41}
\end{equation}

\begin{figure}[hbpt]
\begin{center}
\includegraphics[width=10cm]{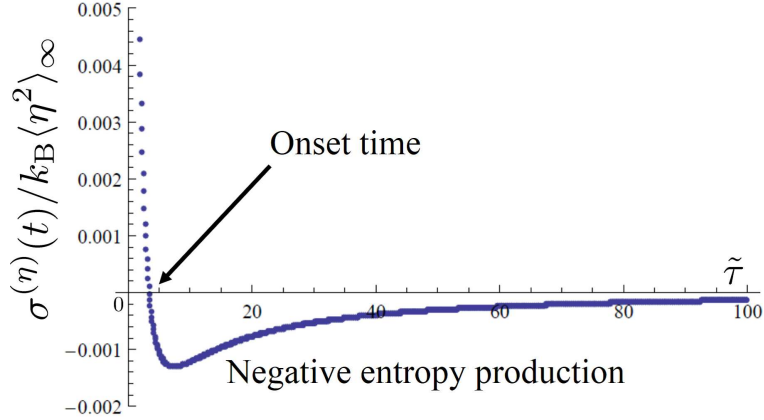}
\end{center}
\caption{Time dependence of the entropy production $\sigma^{(\eta)}(t)$. Plotted points are obtained by the numerical differentiation of the entropy $S^{(\eta)}(t)$ shown in Fig.\ref{fig4}. It is clearly shown that the entropy production becomes negative beyond the onset time.}
\label{fig5}
\end{figure}

Furthermore, the entropy production $\sigma^{(\eta)}$ is obtained by the numerical differentiation of the entropy $S^{(\eta)}(t)$ and it is shown in Fig.\ref{fig5}.
As shown in Figs.\ref{fig4} and \ref{fig5}, the entropy decreases beyond the onset time $t_{\text{o}}^*$.
This effect corresponds to the order formation of our model.
It is interesting to note that the entropy $S^{(\eta)}(t)$ becomes very small as time $t$ increases.
Then the total entropy $S(t)\equiv S^{(m)}(t)+S^{(\eta)}(t)$ increases for certain parameters $a,b$ and $c$ in Eqs.(\ref{eq13}) and (\ref{eq14}).
This is a typical behavior of our system.
When the system includes only one order parameter as shown in the ordinary scaling theory, the total entropy decreases beyond the onset time $t_{\text{o}}$ \cite{1,2}.
On the other hand, when the system includes even and odd order parameters, the ordered state characterized by the odd order parameter does not appear and the total entropy includes the increasing fluctuation $\langle m^2 \rangle$.
Thus, the total entropy $S(t)= S^{(m)}(t)+S^{(\eta)}(t)$ increases despite a decreasing entropy $S^{(\eta)}(t)$ for certain parameters $a,b$ and $c$.

\section{Typical example of the order formation with two parameters}
\subsection{four-body interaction model}
In this section, we show a typical example of the present order formation processes using a four-body interaction model.
As shown in our previous studies \cite{9}, a coplanar spin-pair (spin dipole) ordered state appears in the following four-body interaction model;
\begin{equation}
\mathcal{H}=-\sum_{\text{plaquettes}}J_{ijkl}S_i S_j S_k S_l\equiv \mathcal{H}_{j}+\mathcal{H}'_{j};\quad \mathcal{H}_{j}\equiv -E_{j}S_{j}.\label{eq42}
\end{equation}
Here the Hamiltonian $\mathcal{H}_j$ includes the interactions with spin $S_j$ while the Hamiltonian $\mathcal{H}'_{j}$ does not include the spin $S_j$.
The kinetic theory of the Ising models gives the following relations \cite{12}
\begin{equation}
\tau_{m}\frac{d}{dt}\langle S_j \rangle =- \langle S_j \rangle +\langle \tanh (\beta E_j) \rangle \label{eq43}
\end{equation}
and
\begin{equation}
\tau_{\eta}\frac{d}{dt}\langle S_i S_j \rangle =-2 \langle S_i S_j \rangle + \langle S_i \tanh (\beta E_j) \rangle + \langle S_j \tanh (\beta E_i) \rangle. \label{eq44} 
\end{equation}
The mean-field (decoupling) approximation such as $\langle \tanh (\beta E_j) \rangle \sim \tanh \beta \langle E_j \rangle $ and the correlation identities \cite{14,15} are applied to the above kinetic models (\ref{eq43}) and (\ref{eq44}).
Then we obtain the relations
\begin{align}
\tau_{m}\frac{d}{dt}\langle S_j \rangle = -\langle S_j \rangle + ({\text {higher order terms}}) \label{eq45}
\end{align}
and
\begin{align}
\tau_{\eta}\frac{d}{dt}\langle S_i S_j \rangle &= a(T) \langle S_iS_j \rangle -b(T) \langle S_iS_j \rangle^3 \notag
 \\
&+c(T) \langle S_i \rangle ^2 \langle S_iS_j \rangle + ({\text {higher order terms}}), \label{eq46}  
\end{align}
below the critical temperature.
The thermal noise $\zeta(t)$ is assumed to affect directly the spins.
Finally, using the notation $\eta=\langle S_i S_j \rangle$ and $m=\langle S_i \rangle$, the Langevin equations (\ref{eq13}) and (\ref{eq14}) are obtained for the Hamiltonian (\ref{eq42}).

This is a typical example of the present order formation with two different order parameters.

\subsection{Spin-glass model}
It may be useful to compare the above four-body interaction model with a spin-glass model.
Here we consider the Hamiltonian
\begin{equation}
\mathcal{H}=-\sum_{i,j}J_{ij}S_{i} S_{j},\label{eq47}
\end{equation}
including random interactions $\{ J_{ij} \}$ between Ising spins $S_{i}$ and $S_{j}$.
The random average of the physical parameter $Q(\{ J_{ij} \})$ is denoted as 
\begin{equation}
[Q(\{ J_{ij} \} )] = \int dJ_{ij} P(\{ J_{ij} \} )Q(\{ J_{ij} \}), \label{eq48}
\end{equation}
using the distribution function of $\{ J_{ij} \}$, namely $P(\{ J_{ij} \})$.
As shown in Eq. (\ref{eq43}), the spin moment $\langle S_i \rangle$ follows the equation of motion as
\begin{align}
\tau \frac{d}{dt}\langle S_i \rangle &= -\langle S_{j} \rangle +\left< \tanh \left( \sum_{i,j} K_{ij} S_{j} \right) \right>\notag
\\
&\simeq -\langle S_{j} \rangle +\tanh \left( \sum_{i,j} K_{ij} \langle S_{j} \rangle \right),\label{eq49}
\end{align}
where the parameter $K_{ij}$ is defined as $K_{ij}=\beta J_{ij}$.
Here we use such decoupling approximation as $\langle S_{k} S_{l} \rangle \simeq \langle S_{k} \rangle \langle S_{l} \rangle$.
The present analysis corresponds to the Sherrington-Kirkpatrick model \cite{17} because the decoupling approximation is the same approximation as a mean field theory \cite{12}.
Thus the time dependence of $\langle S_i \rangle^2$ is obtained as
\begin{align}
\tau \frac{d}{dt}\langle S_i \rangle^2 &= 2 \langle S_i \rangle \tau \frac{d}{dt} \langle S_i \rangle \notag
\\
\Leftrightarrow \frac{\tau}{2}\frac{d}{dt}\langle S_i \rangle^2 &= - \langle S_i \rangle^2 + \langle S_i \rangle \tanh \left( \sum_{i,j} K_{ij} \langle S_{j} \rangle \right). \label{eq50}
\end{align}
In the present study, we are interested in the condition $d\langle S_i \rangle/dt \sim 0$, that is, the odd order parameter (magnetization) $m=[\langle S_{i} \rangle]$ evolutes slower than the even order parameter (spin-glass order parameter) $\eta=[\langle S_{i} \rangle^2](=q)$ does.
Then the second term of the right hand side of Eq.(\ref{eq50}) yields
\begin{equation}
\langle S_i \rangle \tanh \left( \sum_{i,j} K_{ij} \langle S_{j} \rangle \right) \simeq \tanh^2 \left( \sum_{i,j} K_{ij} \langle S_{j} \rangle \right),\label{eq51}
\end{equation}
using Eq.(\ref{eq49}).
Expanding the hyperbolic tangent in Eq.(\ref{eq51}), we obtain the relation
\begin{equation}
\frac{\tau}{2}\frac{d}{dt}[\langle S_i \rangle^2] = -[\langle S_{i} \rangle ^2]+\sum_{k,l} [K_{ik}K_{il} \langle S_{k} \rangle \langle S_{l} \rangle] +\cdots. \label{eq52}
\end{equation}
With the use of the symmetric distribution of $\{ J_{ij} \}$ satisfying
\begin{equation}
[K_{ik}K_{il}]=K^2\delta_{k,l} {\text{ and }} [K_{ij}]=0,\label{eq53}
\end{equation}
we introduce the decoupling approximation of the random average in Eq.(\ref{eq52}) as
\begin{align}
&[K_{ik}K_{il} \langle S_{k} \rangle \langle S_{l} \rangle] \sim K^2 q, \label{eq53}
\\
&[K_{ik_1}K_{ik_2}K_{ik_3}K_{ik_4} \langle S_{k_1} \rangle \langle S_{k_2} \rangle \langle S_{k_3} \rangle \langle S_{k_4} \rangle] \sim K^4 q^2,\label{54}
\end{align}
and so on \cite{9}.
Finally, we arrive at the time development of even order parameter $q$ as
\begin{equation}
\tau'\frac{d}{dt}q = (z K^2 -1)q - \frac{2}{3}z^2K^4q^2 + b(T)q^3 + c(T)qm^2 + (\text{higher order terms}), \label{eq55}
\end{equation}
where $\tau'=\tau/2$ and the parameter $z$ denotes $\tau/2$ and the number of nearest neighbor spins.
Here, we have $b(T)\sim K^6$ and $c(T)\sim z^2 K^4$.

As shown in Eq.(\ref{eq55}), the spin-glass model is different from the models described by Eqs.(\ref{eq13}) and (\ref{eq14}) in that the squared term of the order parameter $q$, namely $- 2z^2K^4q^2/3$, exists in the former model.
This is because the spin-glass model belongs to a universality class different from that of the four-body interaction model, as shown in the previous studies \cite{9}.
Actually, this squared term of the spin-glass order parameter $q$ is also derived from the well known free energies of spin-glass models \cite{16,17}. 
Then the properties of the order formation process of spin-glasses are different from our model described by Eqs.(\ref{eq13}) and (\ref{eq14}).
As is seen in this comparison, the order formation process of spin-glasses is more complicated because it includes not only plural order parameters but also the randomness and frustration.

\section{Summary and discussions}
In the present study, we have introduced a simple system phenomenologically including even and odd order parameters and we have analyzed by the Langevin equations, the Fokker-Planck equations, and the non-equilibrium entropies.
As a result, we have clarified the order formation processes from the view point of the parity.
The present system may be useful for understanding the order formation processes of the complex systems characterized by many order parameters.
Especially, the present study indicates a possibility of the increasing total entropy of the complex systems with a certain condition even on the ordered phase.

By the way, it seems to be a little bit strange that the expression of the moment $\langle \eta^2 \rangle$ in Eq.(\ref{eq40}) is different from that in Eq.(\ref{eq15}) for the same physical quantity.
This is because we have made a simple decoupling approximation to obtain the Langevin equation (\ref{eq15}), though a sophisticated approximation has been made to obtain the Fokker-Planck equation (\ref{eq40}).
Since the decoupling approximation corresponds to a kind of mean-field theory, the moment $\langle \eta^2 \rangle$ in Eq.(\ref{eq15}) is different from the fluctuation obtained by the Fokker-Planck equations (\ref{eq18}) and (\ref{eq19}).
In the above reasons, the expressions are different from each other.
However, both of the expressions (\ref{eq15}) and (\ref{eq40}) show the same behavior qualitatively, that is, the initial fluctuation of the {\it even} order parameter is enhanced by the noise through the fluctuation of the {\it odd} order parameter.
This is one of the most important physical results on the order formations in complex systems.

For further study, we will try, in the near future, to clarify the order formation process of spin-glasses.

\bibliographystyle{model1a-num-names}

\end{document}